# Power laws prevail in medical ultrasound

K J Parker[1]

[1]Department of Electrical and Computer Engineering, University of Rochester, 724 Computer Studies Building, Box 270231, Rochester, NY 14627, USA

Email: kevin.parker@rochester.edu

**Abstract**

Major topics in medical ultrasound rest on the physics of wave propagation through tissue. These include fundamental treatments of backscatter, speed of sound, attenuation, and speckle formation. Each topic has developed its own rich history, lexicography, and particular treatments. However, there is ample evidence to suggest that power law relations are operating at a fundamental level in all the basic phenomena related to medical ultrasound. This review paper develops, from literature over the past 60 years, the accumulating theoretical basis and experimental evidence that point to power law behaviors underlying the most important tissue-wave interactions in ultrasound and in shear waves which are now employed in elastography. The common framework of power laws can be useful as a coherent overview of topics, and as a means for improved tissue characterization.

**Keywords:** ultrasound; elastography; power laws; fractals; backscatter; attenuation; dispersion

## 1. INTRODUCTION

Medical ultrasound rests on a robust synthesis of disciplines including acoustics, biophysics, fields and waves, and clinical medicine. As such, the major subtopics in medical ultrasound, including backscatter, attenuation, and speed of sound in soft tissues, can appear as disconnected subfields, each with its own set of principles and grammatical rules. This is amplified by the rich history associated with each topic, augmented by close parallels with historical developments in optics, radar, and sonar. For example, in August of 1880, Lord Rayleigh published an influential paper on the mathematics of the sum of random phases which still applies to speckle from a wide variety of systems including ultrasound and laser imaging (Rayleigh, 1880, 1897, 1918a, b). Rayleigh's pioneering work led to the definitive early textbooks on sound (Rayleigh, 1945a, b), and these along with his study of electromagnetic waves spurred many later developments in scattering. Similar paths can be charted for each major subtopic in medical ultrasound. The proliferation of subtopics increased dramatically with the dawn of tissue characterization beginning in the 1970s when digital recording instruments and computer capabilities made possible the goal of differentiating subtle underlying properties of echoes. A variety of *models* were introduced to provide a rationale for echoes from tissues, along with a proliferation of *parameters* to be estimated, hopefully useful as sensitive measures of normal vs. diseased conditions. In some areas a *dominant paradigm* emerged by subtopic. For example, in shear wave propagation many researchers utilize a Kelvin-Voigt model of tissue rheology. In backscatter, many apply spherical (or modified spherical) theories to model echoes from cells. In speckle studies, many rely on the Rayleigh distribution (1880) or its close relatives developed later in radar or sonar studies. The net result of all this for new graduate students and practitioners in ultrasound is a set of special topics that may seem unrelated in theory yet must be jointly considered in practice. This is an emerging issue for clinicians as well since modern ultrasound scanners are now able to estimate more parameters with a proliferation of units, techniques, and interpretations.

Perhaps there is an underlying framework that can be utilized to everyone's advantage and can bring a sense of commonality to the major subtopics of waves in tissues. The hypothesis set forth herein is that *power laws are fundamental to medical ultrasound, operating at a deep level in each subfield*. Once understood in this framework, the task of tissue characterization may also be simplified with more interconnections between subtopics and parameters.

This review paper highlights key references related to power laws, first as a general topic prevalent in the physical and social worlds, then as specific historical evidence in the primary subtopics of: backscatter, attenuation, speed of sound, speckle, shear wave speed, and shear wave attenuation. Due to constraints on length and in the interest of brevity, we do not review the extensive literature pertaining to each of the dominant paradigms, instead focusing on the collected evidence supporting the power law hypothesis.

## 2. THEORY AND EVIDENCE

### 2.1 The prevalence of power laws in the physical and technological world

At a high level of overview, an important consideration is that power laws appear frequently in different fields, from astronomy to biology to network theory. We define a power law as any relationship between a measured parameter $P_m$ and an independent variable raised to a power, for example, $P_m = f^a$ or $P_m = 1/f^a$ where $f$ is frequency in our later examples and $a$ is the power law parameter or exponent. Close relatives include asymptotic power laws for example $P_m = 1/ (1 + f)^a$, which avoids a singularity at the origin, and $P_m = P_0 + f^a$, which allows for a specified constant value $P_0$ when the independent variable approaches zero. An important property of power laws is that they are evident on log-log plots, i.e., in a plot of log($P_m$) vs log($f$), the simple relation forms a straight line with slope of $a$. In a major review paper by Newman (2005), numerous examples are given of a surprising range of multi-scale phenomenon that follow a power law. From the physical world this includes: the size distribution of craters of the moon, the distribution of solar spots, and distributions of rainfall events and earthquakes. In the human world, power laws also appear in a variety of important areas including income distributions within a country, and the popularity of internet sites. All of these can be thought of as common distributions where there are only a few large-scale items or events, interspersed with many more smaller scale items. This is distinctly different from the other well known "bell curve" or "normal" or Gaussian distribution tending towards a mean value. The power law distribution is also related to a number of well-referenced phenomena known by the Pareto principle, the 80/20 rule, and Zipf's law across economics, business, and actuarial fields (Newman, 2005).

### 2.2 Power laws in tissue structures

Soft tissues are comprised of multiscale features, from subcellular organelles to millimeter scale and centimeter scale compositions. An example of the multiscale structures specific to tendons is given in **Figure 1**, however for each organ a similar diagram could be constructed, from subcellular structures to macroscopically observable features.

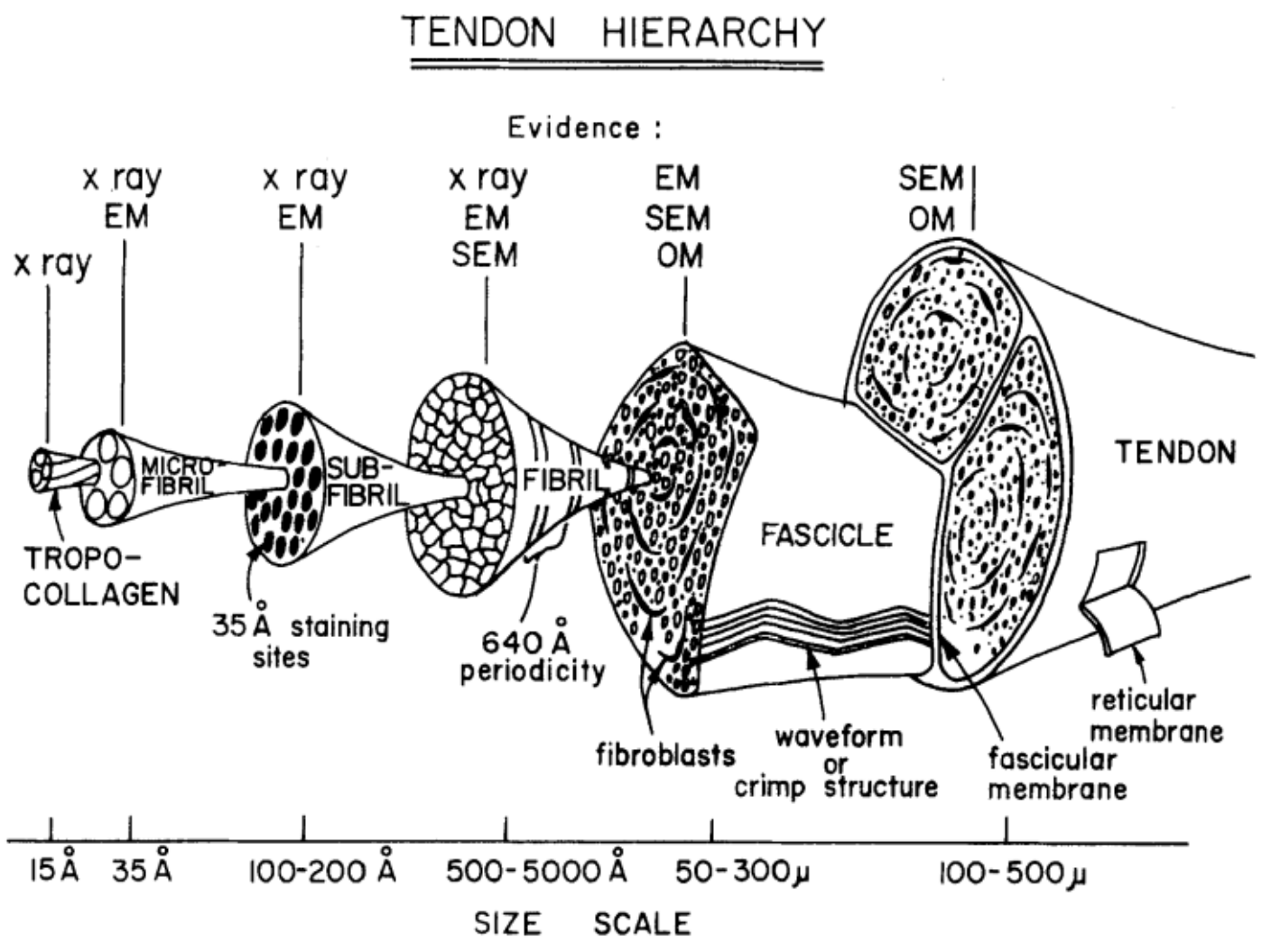

**Figure 1** Hierarchy of structure of a tendon according to Kastelic *et al.* (1978). Evidences are gathered from x-ray, electron microscopy, scanning electron microscopy, and optical microscopy. (Kastelic *et al.*, 1978). Used with permission from Taylor & Francis.

Many structures within the body, particularly the vascular tree and bronchial trees, have been characterized as fractal structures, which are multiscale, space filling, and self-similar over a wide range of sizes (Mandelbrot, 1977; Bassingthwaighte and Bever, 1991; Glenny *et al.*, 1991). These fractal structures' autocorrelation functions are described by power laws (Vicsek, 1992; Carroll-Nellenback *et al.*, 2020). Thus, it should not be surprising that power law relations would propagate into key mathematical models for tissue-ultrasound interactions. Physically, this is consistent with the intuitive idea that there are many small components within organs and cells that influence wave propagation within tissues, along with fewer larger components that also influence the wave propagation, leading to a multiscale phenomenon. The power law models for specific parameters are described in the next sections and summarized in **Table 1**. These apply generally to soft vascularized tissues within typical imaging frequencies (2-20 MHz) and elastographic shear wave frequencies (50-500 Hz) and are not intended (without further considerations) to apply to highly anisotropic structures such as the tendon, muscle, and arteries or extrapolations to ultra-high or - low frequency bands. The references supporting the values and forms shown in **Table 1** are given by subtopic in following sections 2.3 – 2.8.

**Table 1** Power laws governing waves in soft vascularized tissues

| Parameter | Form | Example |
|---|---|---|
| Backscatter | $\sigma(f) = B_1 f^{b_1}$ | $b_1 \cong 1.4$ |
| Attenuation | $\alpha(f) = \alpha_0 f^{a_1}$ | $a_1 \cong 1$<br>$\alpha_0 = 0.05$ Np/cm-MHz |
| Speed of sound | $c_\ell(f) = c_0 + c_1 f^{a_\ell}$ | $c_0 \cong 1500$ m/s |
| Speckle intensity | $P(I) = (b-1)/(1+I)^b$ | $b \cong 3$ |
| Shear wave speed | $c_s(f) = c_{0_s} + c_{1_s} f^{a_s}$<br>OR<br>$c_s(f) = c_{1_s} f^{a_s}$ | $a_s \cong 0.1$ |
| Shear wave attenuation | $\alpha_s(f) = \alpha_{s_0} f^{a_{s1}}$ | $a_s \cong 1$ |

For a complete picture, it should be noted that there has been a rich development of models related to ultrasound propagation in tissues including scattering, attenuation, and speckle that are not tied explicitly to power laws. These include Chivers (1977), Burckhardt (1978), Lizzi *et al*. (1983), Jakeman and Tough (1987), Insana *et al*. (1990), Chen *et al*. (1994), Chen *et al*. (1997), Shankar (2000) Ng *et al*. (2006), Destrempes and Cloutier (2010, 2021), Destrempes *et al*. (2016), Oelze and Mamou (2016). These have been found to match experimental results in a number of settings. However, the following sections 2.3 – 2.8 will focus on the models and experimental evidence tied explicitly to power law behaviors.

## 2.3 Power laws in ultrasound backscatter

A useful model of backscatter vs. frequency $f$, supported by theory and experiments, is given by:

$$\sigma(f) = B_1 f^{b_1}, \tag{1}$$

where $\sigma$ is the differential cross sectional backscatter with units of 1/Sr-cm, and $B_1$ and $b_1$ are the amplitude and power law coefficient, respectively. In theory, it was shown by Bamber (1979) that under a number of different models for random scattering autocorrelation functions, backscatter would have a power law dependency (at least over some limited frequency bandwidth), and the coefficient $b$ could approach 4 at long wavelengths (Rayleigh scattering). Waag *et al*. (1983) demonstrated the tight link between tissue autocorrelation structures (including those inferred from histology slides) and the resulting ultrasound scattering behavior.

Later, Javanaud (1989) explicitly considered a fractal distribution for scatterers in tissues, and showed that a power law dependence on frequency as the expected result from Born scattering formulations. Campell and Waag (1984) demonstrated in theory and experiment that a coefficient of $b_1 = 1.4$ was characteristic for calf liver in the frequency range of 3-7 MHz that is commonly used in abdominal imaging of humans. Other researchers reported similar power law trends (Nicholas, 1982; Landini *et al.*, 1987; Wear *et al.*, 1989; Anderson *et al.*, 2001; Nam *et al.*, 2011).

Recently, by carefully considering the fractal branching vasculature as the primary source of weak (Born approximation) scatterers within the liver illustrated in **Figure 2**, we have been able to experimentally determine the autocorrelation function as a power law and theoretically predict the power law backscatter relationship (Parker *et al.,* 2019a; Parker, 2019a, b; Carroll-Nellenback *et al.*, 2020). A fundamental relationship underlying all of these is that for fractal structures, the spatial autocorrelation function is a power law (Vicsek, 1992). Since backscatter can be related to the autocorrelation function of weak random inhomogeneities (Waag *et al.*, 1983; Morse and Ingard, 1987), when this is introduced into scattering equations these also produce a power law.

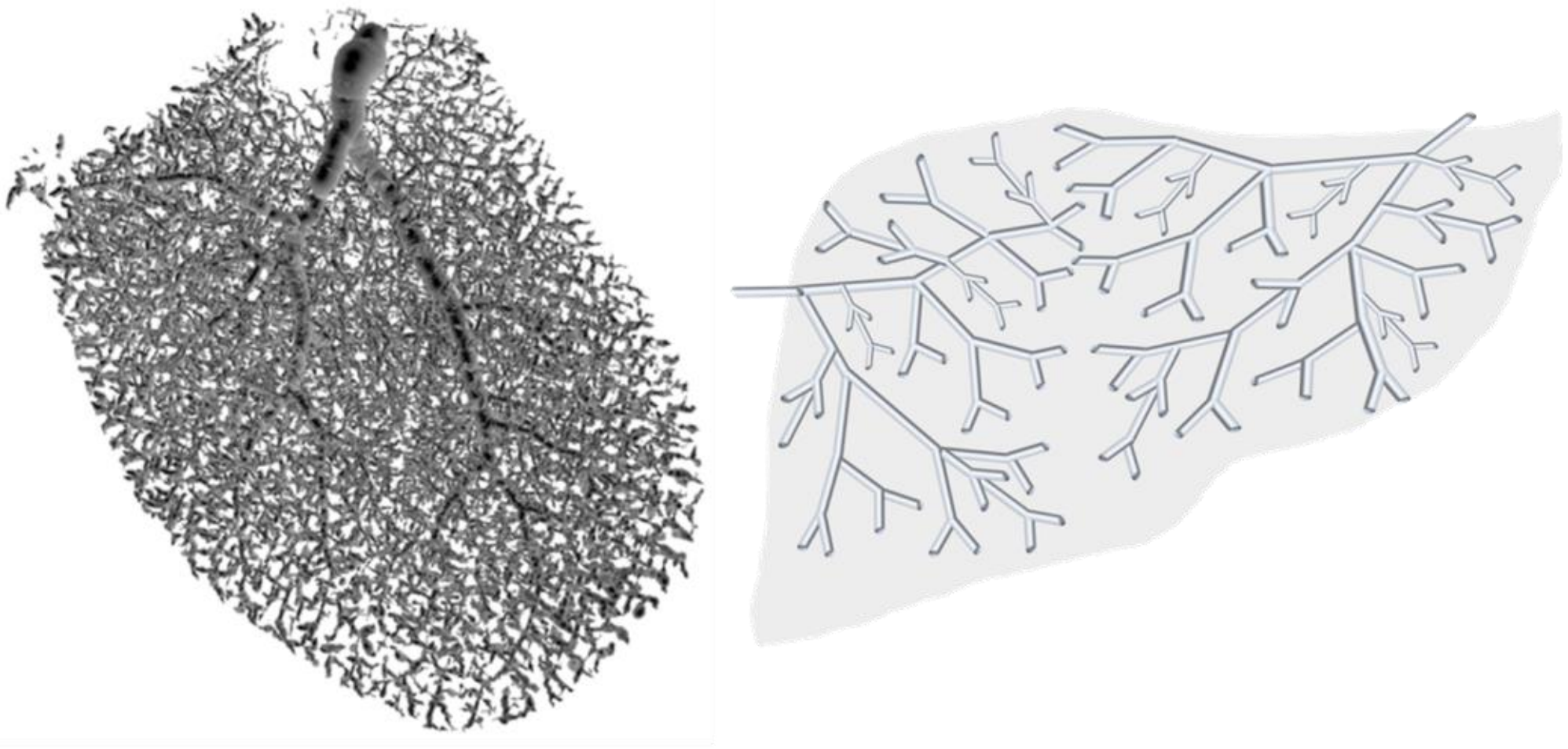

**Figure 2** Fractal branching vasculature (left) from a vascular cast, imaged with micro-CT, and a mathematical model using a power law distribution of cylinders (right) for the study of backscatter. Details of this approach can be found in Carroll-Nellenback *et al*. (2020).

When specific fractal cylindrical models are evaluated, the backscatter vs. frequency exhibits a power law similar to the data collected by Campbell and Waag for bovine liver, as shown in **Figure 3**.

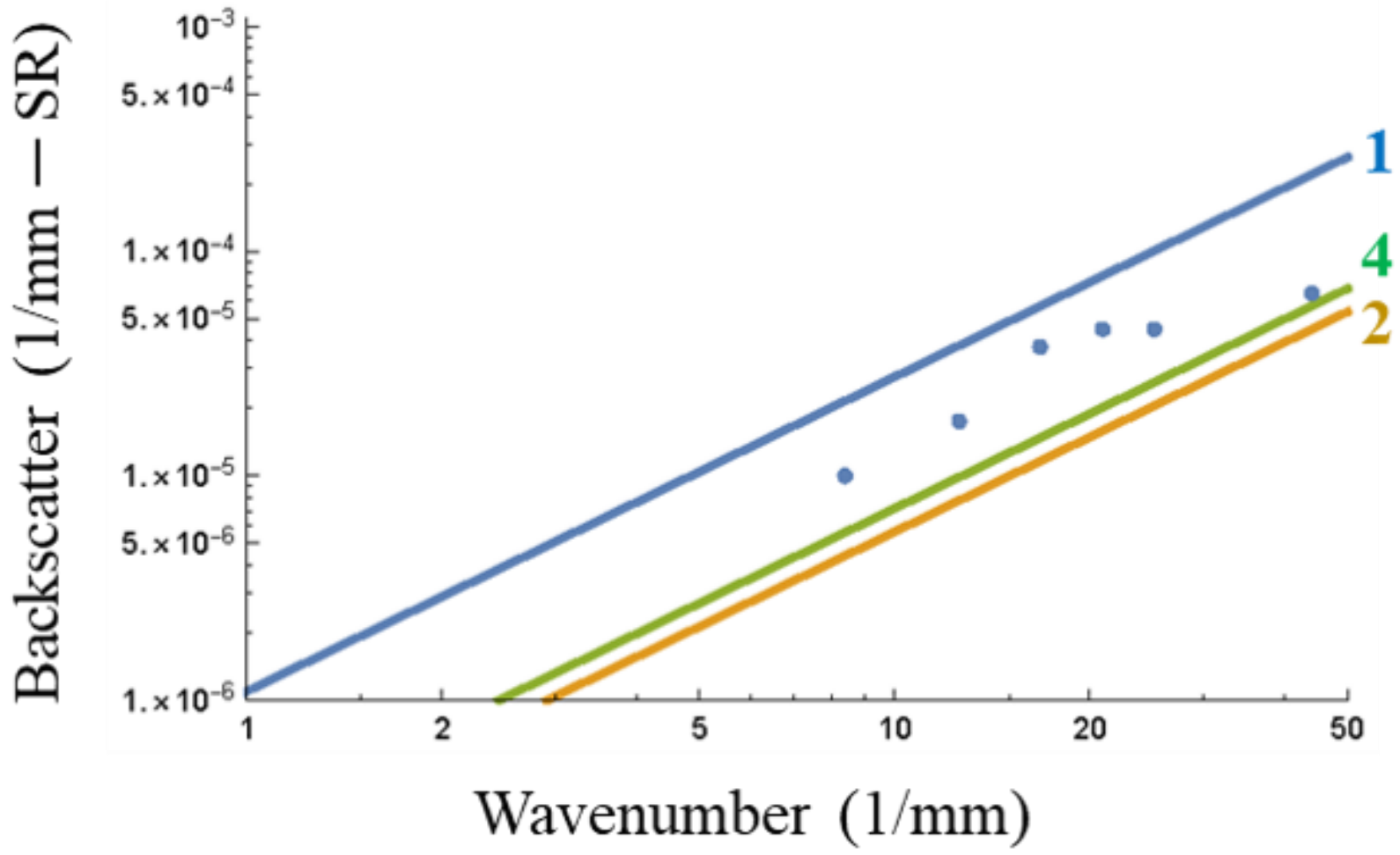

**Figure 3** Theoretical versus experimental backscatter versus wavenumber using values from Campbell and Waag (1984) for *ex vivo* calf liver (dots) as compared with theory. Solid lines are, from upper to lower: Case 1, fluid-filled fractal cylinders as weak scatterers; Case 4, fractal cylinders with a modified Gaussian radial shape; Case 2, fractal cylinders with a parabolic radial shape (Parker, 2019b).

### 2.4 Power laws in ultrasound attenuation

A useful model of ultrasound (longitudinal wave) attenuation vs. frequency, supported by theory and experiments, is given by:

$$\alpha(f)=\alpha_0 f^{a_1}, \tag{2}$$

where $\alpha$ is the attenuation per distance $x$ of a plane wave of frequency $f$, observed as a progressive loss of amplitude during propagation as $\exp[-\alpha \cdot x]$. In acoustic textbooks, attenuation as a function of frequency is usually introduced in the simple form of a relaxation function with a single time constant (Pierce, 1981; Kinsler *et al.*, 1982; Blackstock, 2000) tied to a specific chemical or thermal mechanism. For example, Chapter 9 of Blackstock (2000) discusses relaxation of boric acid and magnesium sulfate in seawater, and general thermal conduction effects, within the treatment of loss mechanisms. Relaxation mechanisms produce an attenuation that has a power law of 2 at low frequencies (compared to 1/time constant). However, tissue components have *multiple* relaxation functions over a *wide range* of characteristic times and strengths. This was seen as early as 1959 by Carstensen and Schwan (1959) studying the absorption of hemoglobin solutions, and even within this particular biomaterial they concluded that "It has been possible to relate quantitatively the magnitude of the absorption and dispersion through relaxation theory by assuming broad distribution of relaxation times." The power law coefficient that best fit their data (0.5 to 10 MHz) was $a_1 = 1$. An early consensus emerged that a power law fit near 1 was adequate for attenuation models of soft tissues (Wells, 1975; Goss *et al.*, 1979; Kuc and Schwartz, 1979; Kuc, 1980; Narayana and Ophir, 1983; Flax *et al.*, 1983; Parker and Waag, 1983; Parker *et al.*, 1984).

At a deeper theoretical level, additional insight has been developed by consideration of the wave equation with loss mechanisms, causality constraints, and multiple relaxation functions leading to fractional derivatives. These have been developed in recent decades by an impressive set of papers by Szabo (1994, 1995, 2003; Szabo and Wu, 2000), Holm (2019; Chen and Holm, 2003; Nasholm and Holm, 2011; Holm and Nasholm, 2014), McGough (Kelly *et al.*, 2008; Kelly and McGough, 2009; Zhao and McGough, 2018), Cobbold (2004; Sushilov and Cobbold, 2003, 2004), and Treeby and Cox (2010, 2011, 2014) . Their work leads to a conclusion that the power law coefficient in many tissues will be close to $a_1 \cong 1$, with some variation related to the fractional derivative or multiple relaxation distribution governing the attenuating medium. For a deeper treatment, see equation (18) of Holm and Nasholm (2014) for the low and intermediate frequency regime and the derivations therein, and equation (5.66) in Holm (2019).

**2.5 Power laws in speed of sound**

A useful model of the speed of sound vs. frequency, supported by theory and experiments, is given by:

$$c_\ell(f) = c_0 + c_1 f^{a_\ell}, \tag{3}$$

where $c_0$ is the low frequency speed and $c_\ell$ and $a_\ell$ are related to the dispersion. In general terms, the speed of sound in a liquid like salt water (and soft tissues are comprised of a high percentage of water with salts and buffers in solution) is determined by the bulk compressibility along with any loss mechanisms affecting the wave. The low frequency bulk modulus and acoustic speed of sound of salt water can be measured using sonar and other techniques, and $c_0$ is roughly 1500 m/s depending on conditions including temperature and exact amounts of salts and proteins in solution (Kinsler *et al.*, 1982). However, in tissue with multiple relaxation mechanisms operating over a wide range of time constants, a fractional derivative model becomes relevant and descriptive and the dispersion term is a power law. For a more in-depth treatment see equation (19) of Holm and Nasholm (2014) and equation (5.67) in Holm (2019) for the low and intermediate frequency regimes, and the derivations leading to those results. The approximation for speed of sound in the low to intermediate frequency range (relative to the reciprocal of the time constants of the relaxation mechanisms) is examined in more detail in the **Appendix**, and some examples of ultrasound dispersion in tissues are given in Bhagat *et al*. (1977) and Marutyan *et al*. (2006).

**2.6 Power laws in ultrasound speckle**

A useful model of the probability $P$ of speckle intensity $I$, from tissue, supported by theory and experiments, is given by:

$$P\left(I\right)=\frac{b-1}{\left(1+I\right)^{b}}, \tag{4}$$

for $b>1$ and $I>0$, and where $b$ is related to the underlying power law of the distribution of multiscale scatterers within the tissue. This recent derivation (Parker, 2019a; Parker and Poul, 2020b; Parker and Poul, 2020a) rests on the presence of two sufficient conditions. First, there is assumed a multiscale distribution of scatterers following a generic power law distribution (few large scatterers, many small ones). Second, there is assumed to be a linear increase in echo intensity from the scatterers as a function of size, at least above some minimum. Once these conditions are satisfied, a simple probability transformation results in the classic Burr distribution for the speckle amplitude, and a modified power law, equation (4), for speckle intensity of the raw (uncompressed) echoes, as illustrated in **Figure 4**.

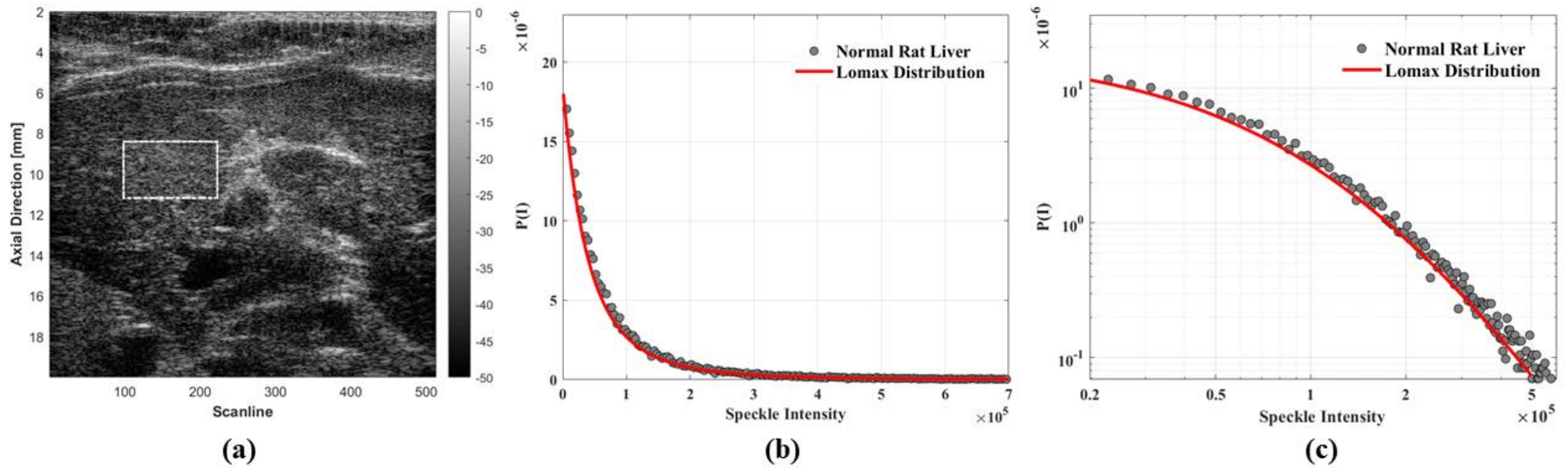

**Figure 4** Speckle patterns in a normal animal liver. (a) B-scan with selected region of interest. (b) Histogram of speckle intensity plotted on a linear scale. (c) Histogram plotted on a log-log scale. The solid red line is the theoretical fit to the asymptotic power law known as the Lomax distribution shown in **Table 1**.

## 2.7 Power laws in shear wave speed

A useful model of the shear wave speed vs. frequency, supported by theory and experiments, is given by:

$$c_s\left(f\right)=c_{0_s}+c_{1_s}f^{a_s}, \tag{5}$$

or in many cases simply:

$$c_s\left(f\right)=c_{1_s}f^{a_s}, \tag{6}$$

where $c_{0_s}$ is the low frequency speed and $c_{1_s}$ and $a_s$ capture the dispersion. In the case of shear wave propagation through tissue, the appropriate theory comes from consideration of transverse waves in a viscoelastic material. A key parameter is the shear modulus along with loss mechanisms associated with viscosity. Of the many models available, the Kelvin-Voigt fractional derivative (KVFD) model captures the multiple relaxation nature of multiscale responses of tissue in shear (Koeller, 1984; Bagley and Torvik, 1986; Zhang *et al.*, 2007; Parker *et al.*, 2019b). The phase velocity of shear waves is directly related to the square root of the shear modulus, so if the shear modulus is governed by a fractional derivative (power law), then so is the shear wave speed, albeit with a power law reduced by ½ due to the square root relationship between the shear modulus and phase velocity (Parker *et al.*, 2018a) and with further details in the **Appendix**. An example of shear wave dispersion is shown in **Figure 5**.

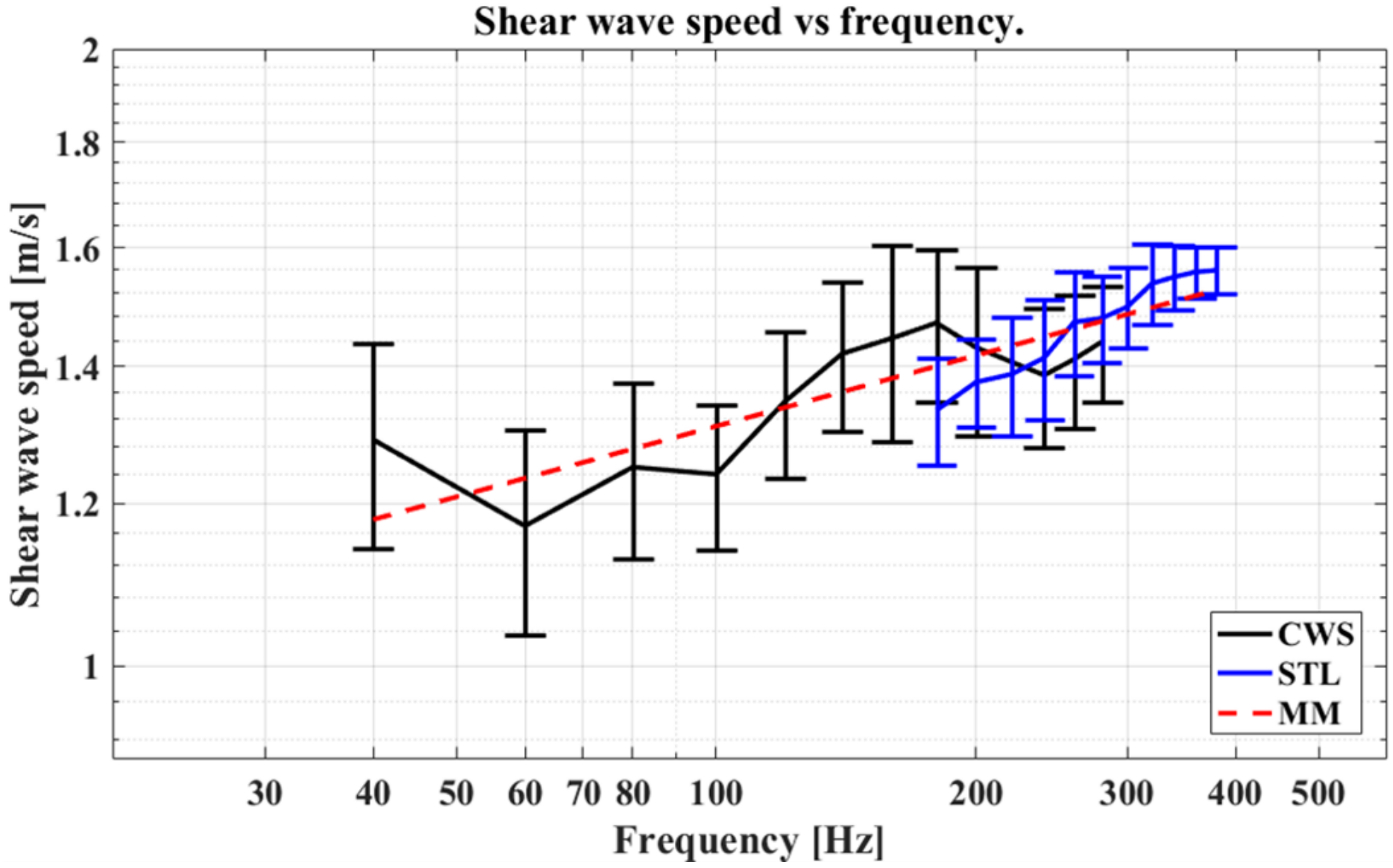

**Figure 5** Shear wave speed $c_s(f)$ data extracted from bovine liver samples, including estimates derived from crawling waves (CWS), single tracking line shear wave estimators (STL), and curve-fit based on stress relaxation results (MM). The nearly linear (on log-log scale) combined results are consistent with the concept of power law behavior (Parker *et al.*, 2019b).

## 2.8 Power laws in shear wave attenuation

A useful model of shear wave attenuation vs. frequency, supported by theory and experiments, is given by:

$$\alpha_s(f) = \alpha_{s_0} f^{\alpha_{s_1}}, \tag{7}$$

where $\alpha_s$ is the attenuation coefficient, and $\alpha_{s_0}$ and $\alpha_{s_1}$ form the power law dependence. Analogous to the discussion and references given in sections 2.4 and 2.7, if the shear modulus has real and imaginary power law dependence on frequency (consistent with a fractional derivative model such as the KVFD model), then the attenuation will also have a power law dependence, close to unity for soft tissues such as the liver (Parker *et al.*, 2015; Nenadic *et al.*, 2017; Parker *et al.*, 2018b; Sharma *et al.*, 2019).

## 2.9 Diagnostic value of power law parameters

Measured values of ultrasound attenuation, backscatter, and speed of sound have been of interest since the early days of tissue characterization (Goss *et al.*, 1978; Linzer, 1979; Bamber and Hill, 1981). Thus, the power law parameters in **Table 1** specific to any tissue or organ will vary with different pathologies and conditions, hence they are important quantities for diagnostic purposes. However, there are over ten of these in **Table 1** and some scanners could estimate additional parameters from contrast studies, nonlinear behaviors, and anisotropic measures. Assuming these can be measured accurately and are independent of system settings and overlying tissues, the net result is a potentially large set of numbers for a clinician to consider along with the traditional B-scan information. To address this realty, some specific approaches have been recently proposed. First, the H-scan analysis estimates the backscatter power law transfer function and attenuation using matched filters (Parker and Baek, 2020). Secondly, a means of combining all available information in a multiparametric space, and then simplifying the result for display purposes has been developed and is termed disease-specific imaging (DSI) (Baek and Parker, 2021). The key idea behind DSI is that any set of parameters listed in **Table 1** along with other available measures will tend to cluster in multidimensional space for a normal organ such as the liver. Furthermore, major disease categories exhibit unique, separable clusters. In this framework any new patient's measures can be compared with known clusters by applying a mathematical measure of closeness in multidimensional (multiparametric) space. This result can be converted into a single metric or a unique, disease-specific color overlay representing the best match of the new patient's data to the closest known cluster of disease states, as shown in **Figure 6**. Thus, a clinician using an advanced ultrasound scanner does not need to memorize the physics and normal values associated with backscatter, speed of sound, and so forth. Instead, the DSI analysis in multiparametric space will identify the closest likely match, visually and quantitatively.

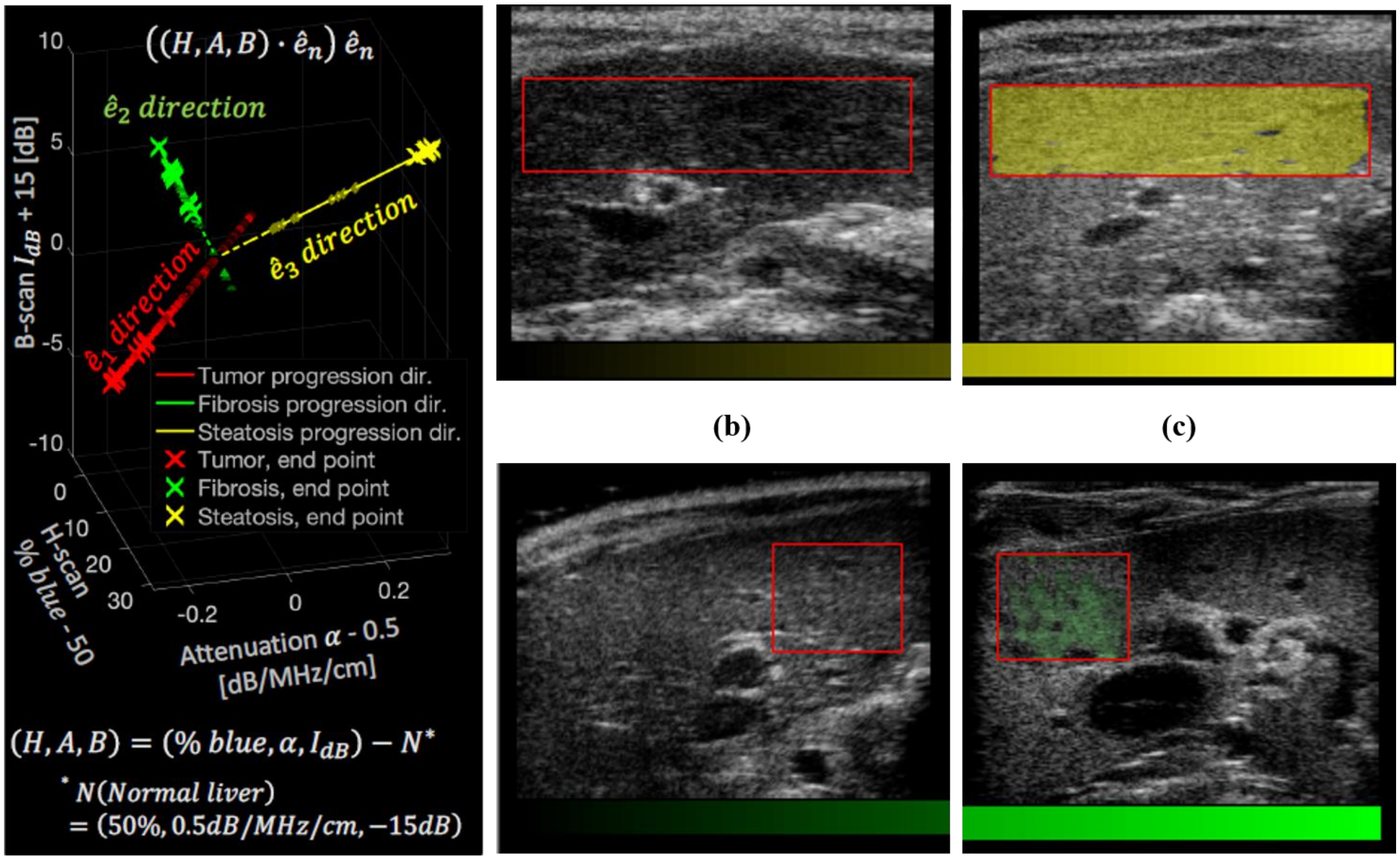

**(a)** **(d)** **(e)**

**Figure 6** Disease-specific imaging. (a) Three features of formal and diseased liver tissues were extracted by H-scan, attenuation, and B-scan analyses, generating the components ($\%\ blue$, $\alpha$, $I_{dB}$), respectively. The coordinate represents a normalized (*H*,*A*,*B*) = ($\%\ blue$, $\alpha$, $I_{dB}$) – (50,0.5,-15) for convenience, representing changes from normal liver (50%, 0.5 dB/MHz/cm, -15 dB). Linear lines show disease progression) pathways from normal to end time point causing death. All measured features of this study were projected onto their diseases' pathway, showing scatter plots. The end points causing death are denoted using the symbol "×". Then the distance between the origin and any new measurement can be applied as color intensity, meaning the greater values represent more severe disease. The support vector machine classifies liver states, and assigns colors of yellow and green for steatosis and fibrosis, respectively. Then the colors are overlaid on B-scan (b-c, d-e). Normals in (b, d) only show B-scan without colors. Late stages in (c,e) fill in with saturated colors, representing extensive disease. (Baek and Parker, 2021). Used with permission from John Wiley and Sons.

## 3. DISCUSSION AND CONCLUSION

A number of developments in the past 60+ years have pointed to power law relations in the major phenomena that form the elements of medical imaging. These appear to emerge out of the nature of soft tissue as a complicated multiscale structure with multiple levels of influence on propagating waves. It seems that once a power law is introduced into a tissue model, be it scattering or attenuation or others, the power law propagates through the relevant equations and then will be evident in the properties of echoes from soft tissues. Within this framework the task of tissue characterization immediately turns to the estimation of key parameters listed in **Table 1**.

Furthermore, there are additional parameters that can be measured and included in a comprehensive assessment of a tissue. For example, in the classical “weak shock theory” of nonlinear wave propagation, the $n^{th}$ harmonic generated by shock formation of an initially pure tone wave will have amplitude $B_n \sim 1/n$ at long distance and high shock parameter (Blackstock, 1966). This is yet another power law relationship which could be incorporated into a framework that includes **Table 1**. Others will likely be added as more types of measures are developed.

It should be noted however that linear approximations to a function over small ranges may sometimes be a simple alternative to a power law formula, since the Taylor series approximation, consisting of a constant and a first order term is generally useful for well-behaved functions. So for example, equation (3) under a Taylor series expansion would have a constant term plus a first order term, $f^1$ rather than $f^{a_1}$. This seems confusing, since we then appear to have two different power law exponents describing one set of measurements. However, this is understandable within the confines of the first order Taylor series as a general approximation valid only over a limited range, whereas the power law fit incorporates an *a priori* model tied to multiscale distributions and the power law behavior they exhibit.

In any organ there is, of course, an upper and lower limit on the size of structures that might contribute to scattering, so that implies an upper and lower limit on the range of applicability of equation (1), and similarly for all of the power law parameters that are considered. These limits are not well characterized and require further study. For example, Herthum *et al*. (2021) have estimated a possible sharp decrease in shear wave speed near 1 Hz in brain tissue, however the implications of this are not well understood. The low frequency regions represent a challenge for shear wave experimentation because the wavelengths become large compared to adult organs. At high ultrasound frequencies (above 20 MHz) and shear wave frequencies (above 500 Hz) the penetration depths due to increased attenuation can create practical limitations, so these upper limits will also require additional study. Another limitation of the power law framework and forms of **Table 1** is that these are not intended for monodispersed scatterers, periodic structures, or single relaxation materials. While these conditions do not apply to most soft tissues, they can be created in phantoms and patterned cell cultures, so alternative models would be required for these special situations. In this context, it is important to note that models of cells as spheres can also exhibit power laws under certain conditions, especially given randomization of spacing or sizes. For a particular example, the modeling of red blood cell clustering in Savery and Cloutier (2001) introduced aggregation parameters which could be linked to a fractal dimension of the ensemble. In certain cases, this can lead to power law backscatter behavior over many decades of frequencies used in medical ultrasound. This points to the possibility that other earlier models, not explicitly concerned with power laws, may exhibit power law behaviors under particular regimes, however this remains for future research.

Ultimately, it is hoped that a power law framework might simplify the overview of the major subtopics in medical ultrasound. Furthermore, the power law framework can focus tissue characterization on key parameters. These parameters may be ultimately integrated into an improved diagnostic classification by methods such as the multiparametric DSI, however additional work is required to assess the full scope and accuracy of this approach.

**Acknowledgments**

This work was supported by National Institutes of Health grants R21EB025290 and R21AG070331. We gratefully acknowledge the contributions of J.Ormachea, S. Poul, and J. Baek to this work, along with helpful and insightful comments from Professor S. Holm.

**Appendix**

How are the power law relations for wave speed and attenuation generated? One straightforward approach begins with the *general* equations for wave propagation, into which the *specific* behavior of the medium, for example a KVFD model, is inserted. As an example of this approach, start with the complex exponential representation of a plane wave at frequency $\omega$, the monochromatic wave equation for a disturbance traveling in the $+x$ direction is:

$$u(x) = A\mathbf{e}^{-j(kx-\omega t)}, \tag{8}$$

where $u(x)$ is displacement, $A$ is the amplitude, $j$ is the imaginary unit, and $k$ is the wavenumber. Furthermore,

$$k = \frac{\omega}{c_s} \tag{9}$$

and

$$c_s = \sqrt{\frac{G}{\rho}} \cong \sqrt{\frac{E}{3\rho}}, \tag{10}$$

where $c_s$ is the shear wave speed, $\rho$ is density, $G$ the shear modulus, and $E$ the Young's modulus, with the approximation $G = E / 3$ valid for nearly incompressible materials. More details can be found in Chapter 5 of Graff (1975), or reviewed for elastography in Parker *et al.* (2011). In this purely elastic, lossless propagation, there is only one velocity $c_s$ for all frequencies and no attenuation.

However, when some loss mechanisms are present, the speed $c$ changes with frequency (called "dispersion"). In this case, the solution to the lossy wave equation still resembles equation (8), but now $k$ is complex: its real component is related to $\omega / c_p$ where $c_p$ is the phase velocity of the shear wave (which is dependent on frequency), and the imaginary part of $k$ defines an exponential decay with distance. We can introduce this in a general way by defining a complex, frequency-dependent shear modulus (Lakes, 1999; Zhang and Holm, 2016) for the material or tissue:

$$G^*(\omega) = \left(G_d(\omega) + jG_i(\omega)\right), \tag{11}$$

where $G^*$ is the complex modulus, $G_d$ is the dynamic modulus, and $G_i$ is the loss modulus. Introducing this into the form of equations (8), (9), and (10), the wave number is now written as:

$$k = \frac{\omega}{\sqrt{\frac{G_d(\omega) + jG_i(\omega)}{\rho}}} = \beta - j\alpha = \frac{\omega}{c_p} - j\alpha, \tag{12}$$

and the wave propagation now has the form:

$$u(x) = A\mathbf{e}^{-\alpha x}\mathbf{e}^{j(\omega t - \beta x)}. \tag{13}$$

Working through the real and imaginary parts of equation (12) and denoting

$$|G| = \sqrt{G_d^2(\omega) + G_i^2(\omega)}, \tag{14}$$

the result for the phase velocity as a function of frequency is:

$$c_p = \left[\sqrt{\frac{|G|}{\rho}}\right]\left[\frac{1}{2}\left(1 + \frac{G_d(\omega)}{|G|}\right)\right]^{-1/2}, \tag{15}$$

and the attenuation coefficient includes a leading term directly proportional to the first power of frequency:

$$\alpha = \left[\omega\sqrt{\frac{\rho}{|G|}}\right]\left[\frac{1}{2}\left(1 - \frac{G_d(\omega)}{|G|}\right)\right]^{1/2}. \tag{16}$$

The above equations are all general; they are formulated from basic relationships. Particular forms of these depend on the model chosen for $G^*(\omega)$, and examples from some simple tissue models can be found in Carstensen and Parker (2014). However, in this manuscript, we are concerned with power law relations and so as an example, the specific rheological KVFD model can be used for $G(\omega)$ and entered into the above equations for $c$ and $\alpha$:

$$G^*(\omega) = G_0 + G_1(i\omega)^a. \tag{17}$$

The result is somewhat complicated because of the square root of the real and imaginary parts of $G^*$, but simplifies for low frequencies and higher frequencies, where $G_0$ is dominant, or negligible, respectively. We believe that for shear waves around 100 Hz in elastography, the $G_0$ term will be negligible (Zhang *et al.*, 2007), and the terms simplify to:

$$c_p = \omega^{\frac{a}{2}} c_1 \tag{18}$$

and

$$\alpha = \frac{\omega^{\left(1-\frac{a}{2}\right)}}{c_1}\left(\frac{\pi a}{8}\right), \tag{19}$$

where $c_1$ is the phase velocity measured at $\omega = 1$ rad/s, and a small value of $a$ is assumed in equation (19).

This is consistent with equations (5.66) and (5.67) in Holm (2019) for the intermediate frequencies. In any case, we can observe that the square root operator on $G$ plays an important role in the power law observed by introducing a factor of ½ in the speed and attenuation exponents, i.e., the power $a$ in the shear modulus becomes $a/2$ in the speed and attenuation equations.

**ORCID**

K J Parker https://orcid.org/0000-0002-6313-6605